\documentclass[12pts]{article}
\usepackage{epsfig}
\usepackage{fullpage}

\def\be{\begin{equation}}
\def\ee{\end{equation}}

\def\bae{\begin{eqnarray}}
\def\eae{\end{eqnarray}}
\def\deca{Z \to \bar \nu {\nu} \gamma}
\def\decb{Z \to \bar \nu {\nu} \gamma \gamma}

\begin{document}

\title{New physics effects in rare $Z$ decays}

\author{M. A. P\'erez\\Departamento de F\'\i sica, Cinvestav,\\
Apartado Postal 14--740, 07000, M\'exico D. F.
M\'exico.\\mperez@fis.cinvestav.mx\\ \\G.
Tavares--Velasco\\Instituto de F\'isica y Matem\'aticas,\\
Universidad Michoacana de San Nicol\'as de Hidalgo,\\ Apartado
Postal 2--82, 58040, Morelia, Michoac\'an,
M\'exico.\\gtv@itzel.ifm.umich.mx\\ \\J. J. Toscano\\Facultad de
Ciencias F\'\i sico Matem\' aticas,\\ Benem\' erita Universidad
Aut\' onoma de Puebla,\\ Apartado Postal 1152, Puebla, Puebla, M\'
exico.\\jtoscano@fcfm.buap.mx}

\maketitle

\begin{abstract}
Virtual effects induced by new physics in rare $Z$ decays are
reviewed. Since the expected sensitivity of the giga--$Z$ linear
collider is of the order of 10$^{-8}$, we emphasize the importance
of any new physics effect that gives a prediction above this
limit. It is also pointed out that an improvement on the known
experimental constraints on rare $Z$ decays will provide us with a
critical test of the validity of the standard model at the loop
level.
\end{abstract}

\section{Introduction}

Processes that are forbidden or highly suppressed constitute a
natural framework to test any new physics lying beyond the
standard model (SM). In particular, rare $Z$ decays have been
studied extensively in order to yield information on new physics
 \cite{Glover-1}. The major decays of the $Z$ boson into fermion
pairs are by now well established within an accuracy of one part
in ten thousands. While the sensitivity of the measurement for the
branching ratios of rare $Z$ decays reached at LEP--2 is about
$10^{-5}$  \cite{PDG}, future linear colliders (NLC, TESLA) will
bring this sensitivity up to the $10^{-8}$ level  \cite{Aguilar}.
As a consequence, the interest in the study of rare $Z$ decays is
expected to increase.

The general aim of the present paper is to review rare $Z$ decays
that may be induced in the SM at the loop level or by any other
mean from new degrees of freedom predicted by some extensions of
the SM. Since the energy scale $\Lambda$ associated with new
degrees of freedom should be large as compared to the electroweak
scale, it is expected that their virtual effects may show up in
some rare $Z$ decays where the conventional SM radiative
corrections are suppressed. We will not consider thus $Z$ decays
into four fermions as they are induced at the tree level by the SM
$ZWW$ and $ZZh$ couplings  \cite{Glover-2}. For the purpose of the
present paper, we will classify rare $Z$ decays into the following
groups:

\begin{enumerate}
\item Single--photon decays. \item Two--photon decays. \item
Decays with photons and gluons. \item Decays involving scalars.
\item Flavor changing decays, $Z \to l_i^\pm l_j^\mp$, $q_i^\pm
q_j^\mp$.
\end{enumerate}

Our presentation will proceed by addressing each one of the above
sets of channels in the following sections. We will be interested
in comparing SM predictions with those obtained in some of its
extensions. We will find that in most cases the language of
effective theories is the most efficient tool to make an objective
comparison between the known experimental bounds on rare $Z$
decays and the various extension of the SM.

The expression for the total $Z$ decay width $\Gamma_Z$ is
required in the calculation of the branching ratios of rare $Z$
decays. According to our present knowledge of the $Z$ properties
 \cite{PDG}, $\Gamma_Z$ is obtained to a very good approximation by
summing over all the partial decay widths into fermion pairs

\be
\Gamma_Z=\sum_{f\neq t}\Gamma_{f\bar f},
\ee

\noindent where the partial decay widths $\Gamma_{f\bar f}$ are
given in the SM, including QCD and electroweak radiative
corrections, by the following expression  \cite{Baur}

\bae \Gamma_{f\bar f}&=&N_f^c \,\Gamma_0
\frac{\sqrt{1-4\mu_f}}{1+Re(\hat\Pi^Z(m_Z^2))}\left((1+2\mu_f)|g_V^{Zf}(m_Z^2)|^2+
(1-4\mu_f)|g_A^{Zf}(m_Z^2)|^2\right)\nonumber\\
&\times&(1+\delta^f_{QED})(1+\frac{N_c^f-1}{2}\delta_{QCD}),
\eae

\noindent where $N_f=1\,(3)$, for $f=l\,(q)$, is the color factor,
$\Gamma_0=\alpha\, m_Z/3$, $\mu_f=m_f^2/m_Z^2$, $g^{Zf}_{A,V}$ are
the effective coupling constants, $\hat\Pi^Z$ is the $Z$ wave
function renormalization contribution, and the QED and QCD
corrections are given by  \cite{Baur}

\be \delta_{QED}^f=\frac{3\alpha Q_f^2}{4\pi},\ee

\be
\delta_{QCD}^f=\frac{\alpha_s(m_Z^2)}{\pi}+1.405\left(\frac{\alpha_s(m_Z^2)}{\pi}\right)^2
-12.8\left(\frac{\alpha_s(m_Z^2)}{\pi}\right)^3-\frac{Q_f^2\alpha\alpha_s(m_Z^2)}{4\pi^2}.
\ee

We will now proceed to discuss the most interesting rare $Z$
decays.

\section{Single--photon decays}

The single--photon decays $Z  \to  X + \gamma$, where $X$ stands
for any neutral, invisible state, has played a privileged role in
our quest for new physics beyond the SM. The L3 and DELPHI
Collaborations searched for energetic single--photon events near
the $Z$ pole at the CERN LEP collider and set the bound
 \cite{L3-1}

\be BR(Z \to \bar\nu\nu\gamma)\leq 10^{-6}.\label{BRZnng}\ee

\begin{figure}
\begin{center}
\epsfig{file=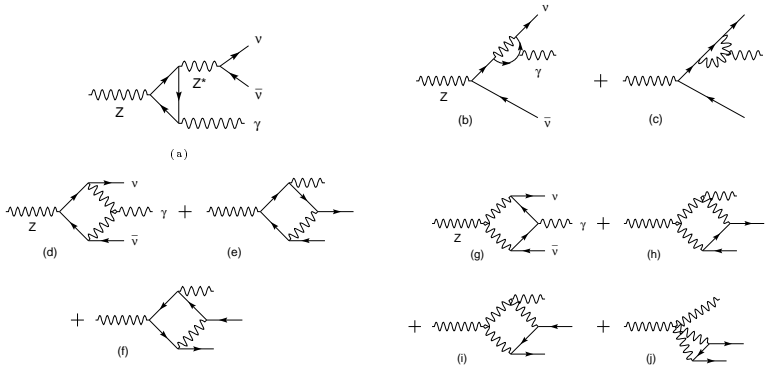,width=4in} \caption{Feynman diagrams
contributing to the decay $\deca$ in the SM {\protect
\cite{Hernandez-1}}. Crossed diagrams must be added.}
\end{center}
\label{ztnngSMfd}
\end{figure}

In the SM this decay is negligibly small and receives
contributions from the Feynman diagrams shown in Fig.
\ref{ztnngSMfd}. It has been found  \cite{Hernandez-1} that the
main contribution comes from a $U_e(1)$ gauge structure induced by
the neutrino magnetic dipole transition (Fig.
\ref{ztnngSMfd}b--\ref{ztnngSMfd}c) and the box diagrams (Fig.
\ref{ztnngSMfd}d--\ref{ztnngSMfd}i). The calculation was performed
in a nonlinear $R_\xi$--gauge, and the result obtained for the
branching ratio is  \cite{Hernandez-1}

\be BR_{SM}(Z \to \bar\nu\nu\gamma)= 7.16 \times 10^{-10}, \ee

\noindent which is about four orders of magnitude below the
experimental limit (\ref{BRZnng}) and thus it leaves open a window
to search for new physics effects in single--photon decays of the
$Z$ boson.

In the early days of the Higgs--boson hunting, when the mass limit
was well below $m_Z$, the decay mode $Z \to h\gamma$ was taken as
one of the best candidates to discover a light Higgs boson. Its
branching ratio in the SM was found somewhat small  \cite{Cahn},
of the order of $10^{-4}$ for $2/3\,m_Z<m_h<m_Z$, but it was
widely studied in various extensions of the SM due to its
sensitivity to gauge couplings and clean signature. For instance,
while in Left--Right (LR) symmetric models this decay mode has
essentially the same width as in the SM  \cite{Martinez}, one
could expect an enhancement of about one order of magnitude in
supersymmetric models \cite{Bates} and in effective theories where
the tree--level generated bosonic operators of dimension 8 become
important  \cite{Hernandez-2}.

The decays of the $Z$ boson into a single--photon plus a virtual
Higgs boson are highly suppressed unless there is a resonant
effect for $m_h<m_Z$. Such was the case for a very light axion
$A$, with a mass $m_A\sim 1$ MeV, with the resonant sequence $Z
\to A\gamma$, $A \to \gamma\gamma$ giving a spectacular event with
three photons  \cite{Kim}. The $ZA\gamma$ and $A\gamma\gamma$
couplings arise from the effective interaction

\be {\cal L}_{Eff.}=\frac{1}{32\pi^2}\frac{A}{F_A}\left(g_c^2
F_{\mu\nu}^a\tilde{F}^{a\ \mu\nu} +e^2
C_{a\gamma\gamma}F_{\mu\nu}\tilde{F}^{\mu\nu}+\frac{2e^2}{s_W
c_W}C_{a Z\gamma}Z_{\mu\nu}\tilde{F}^{\mu\nu}\right),
\label{ZageffL}\ee

\noindent where ${F^a}_{\mu\nu}$, $F_{\mu\nu}$, $Z_{\mu\nu}$ are
the field strengths of the gluon, photon and $Z$ boson,
respectively. For a variant axion with a decay constant $F_A$ of
the order of 10 GeV, this decay mode was estimated to be viable in
the LEP--1 run  \cite{Kim}. However, the chance of having such a
light axion seems to be ruled out  \cite{PDG}.

The possibility of using the rare decay modes $Z \to h\gamma$,
$hh\gamma$, $\bar\nu \nu\gamma$, and $\bar\nu\nu h$ was also
considered in order to test strongly--coupled standard (SCS)
models with a light Higgs boson  \cite{Dusedau}. In this type of
models, the $SU_L(2)$ gauge group is not spontaneously broken but
instead confining. As a consequence, the left--handed quarks and
leptons are fermion--boson bound states and the Higgs and
intermediate vector bosons are boson--boson bound states. The
above decay modes are induced at the loop level but they may be
studied in a model independent way with an effective Lagrangian
similar to (\ref{ZageffL}) for the $Zh\gamma$, $Zhh\gamma$,
$Z\nu\bar\nu\gamma$, and $Zh\nu\bar\nu$ couplings. The conclusion
of this analysis indicated that the decay modes $Z \to h\gamma$,
$\nu\bar\nu\gamma$ were specially suited to test a SCS model with
a light Higgs boson  \cite{Dusedau}. However, this possibility has
been excluded since the lower bound on the Higgs boson mass is
well above the $Z$ boson mass  \cite{Talks}.

In a similar way, single--photon decays were proposed to test the
existence of other very light invisible particles predicted in
some supersymmetric (SUSY) mo\-dels: $Z \to {\widetilde G}
{\widetilde Z} \to {\widetilde G}{\widetilde G} \gamma$, $Z \to
J\gamma$, $JJ\gamma$, where ${\widetilde G}$ is a superlight
gravitino $m_{\widetilde G}\le 10^{-1}$ eV, ${\widetilde Z}$ is
the lightest neutralino  \cite{Dicus}, and $J$ is a (nearly)
massless pseudoscalar Goldstone boson (a Majoron) which appears in
SUSY models with spontaneous violation of $R$--parity
 \cite{Romao-1}, or with spontaneous lepton number violation
 \cite{Romao-2}. In all these decay modes, the respective branching
ratios were in principle accessible to the LEP--1 sensitivity.
Therefore, the negative search for single--photon events in $Z$
decays  \cite{L3-1} can be translated into severe constraints on
the parameters involved in these new physics effects. Even more,
the decay mode into a neutralino and a gravitino is almost
excluded by the current limits on the neutralino mass obtained at
the Tevatron  \cite{PDG}.

 It was also realized long time ago
that the production of single photons at LEP--1 energies
constitute a process which is most sensitive to anomalous
$ZZ\gamma$ couplings due to the large branching ratio for the $Z
\to \bar\nu\nu$ mode and the absence of background from final
state radiation  \cite{Aihara,Ellison}. Since the L3 and DELPHI
Collaborations found that the level of energetic single--photon
events is consistent with what is expected in the SM \cite{Groom},
from the limit (\ref{BRZnng}) it is possible to derive upper
bounds on the $ZZ\gamma$ coupling. The self couplings of photons
and the $Z$ boson constitute the most direct consequence of the
$SU(2)\times U(1)$ gauge symmetry. In the SM they vanish a tree
level and one--loop effects are of the order of $10^{-10}$
\cite{Larios-1}. These couplings have not been measured with good
precision  \cite{Aihara,Ellison,Groom} and any deviation from the
SM prediction may be thus associated with physics beyond the SM.

\begin{figure}
\begin{center}
\epsfig{file=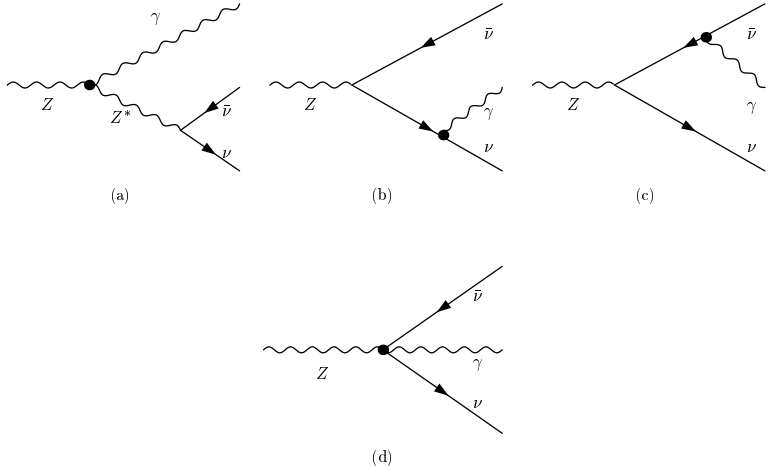,width=4in} \caption{Feynman diagrams with
anomalous couplings contributing to the decay $\deca$ in effective
theories.} \label{ztnngELfd}
\end{center}
\end{figure}

Single photon events coming from $Z$ decays are best analyzed with
the machinery of the effective Lagrangian approach (ELA). The
Feynman diagrams associated with the effective couplings which may
induce the rare decay $Z \to \bar\nu\nu\gamma$ are depicted in
Fig. \ref{ztnngELfd}. They correspond to the contributions
generated by the effective couplings $ZZ^*\gamma$,
$\bar\nu\nu^*\gamma$ and $\bar\nu\nu Z\gamma$. The upper limit
(\ref{BRZnng}) obtained by the L3 and DELPHI Collaborations can be
translated in turn into constraints on these couplings. The
$ZZ^*\gamma$ effective vertex can be parametrized in terms of four
form factors $h_i^Z$. Two of them are CP--conserving and two are
CP--violating  \cite{Larios-1,Maya}

\bae \label{zzgpar} \Gamma_{\alpha \,\beta
\,\mu}^{ZZ^*\gamma}(p,q,k)&=&\frac{ie }{m_Z^2}\bigg[h_{1}^Z
\left(k^{\alpha} g^{\mu \beta}-k^{\beta} g^{\mu \alpha} \right)
 + \frac{h_{2}^Z}{m_Z^2}q^{\alpha}\left(q\cdot k \, g^
{\beta \mu} -k^ {\beta} q^{\mu}\right) \\\nonumber &+&h_{3}^Z
\epsilon^{\alpha \beta \mu \nu} k_{\nu}+
\frac{h_{4}^Z}{m_Z^2}q^{\alpha}\epsilon^{\beta \mu \nu \rho}
p_{\nu} q_{\rho} \bigg] \left(q^2-m_Z^2 \right). \eae

The $\deca$ decay width receives the following contribution from
these vertices  \cite{Larios-1}

\be BR(\deca)= \frac{2g^2}{c_W s_W}2.912\times 10^{-5}
\left(|h^Z_{1}|^2+|h^Z_{3}|^2\right). \ee

\noindent In obtaining this expression, only the CP--conserving
terms were considered as the CP--violating ones are expected to be
strongly suppressed. The experimental bound (\ref{BRZnng}) induces
then the limits

\be |h^Z_{1,\,3}|<0.38, \ee

\noindent which agrees with previous bounds obtained from
scattering experiments  \cite{Groom}.

The LEP--1 bound (\ref{BRZnng}) on the $Z \to \bar\nu\nu\gamma$
decay has also been used to put a direct limit on the magnetic
moment of the $\tau$ neutrino
 \cite{L3-1,Larios-1,Maya,Maltoni,Aydin,Larios-2}. The study of the
neutrino electromagnetic properties have renewed interest since
they may play a key role in elucidating the solar neutrino puzzle
 \cite{Cisneros}: it can be explained by a large neutrino magnetic
moment in the range $10^{-12}\,\, \mu_B$, where $\mu_B$ stands for
the Bohr magneton $ \mu_B=e/(2m_e)$. In the simplest extension of
the SM with massive neutrinos, one--loop radiative corrections
generate a magnetic moment proportional to the neutrino mass
 \cite{Marciano}

\be \mu_\nu=\frac{3G_F\, e\,m_\nu}{8\sqrt{2}\pi}=3.2 \times
10^{-19}\,\left(\frac{m_\nu}{1\,\,\rm\,eV}\right), \ee

\noindent which seems to be too small if one uses neutrino masses
compatible with the mass square differences needed by atmospheric
 \cite{SK}, solar  \cite{SNO}, and the LSND data  \cite{LSND}.

The transition magnetic moments of Dirac neutrinos can be
parametrized through the effective interaction  \cite{Larios-2}

\be {\cal
L}_{\nu_i\bar\nu_j\gamma}=\frac{1}{2}\mu_{\nu_i\bar\nu_j}\bar\nu_i\sigma_{\mu\nu}\nu_j
F^{\mu\nu}, \ee

\noindent where $\mu_{\nu_i}=\mu_{\nu_i\nu_i}$ will correspond to
the $\nu_i$ (diagonal) magnetic moment. Within the effective
Lagrangian framework, the effective $\bar\nu\nu\gamma$ interaction
will induce a contribution given by the Feynman diagram shown in
Fig \ref{ztnngELfd}b. The LEP--1 limit (\ref{BRZnng}) gives the
following bound on the $\nu_{\tau}$ magnetic moment
 \cite{Maya,Larios-2}

\be \mu_{\nu_\tau}<2.62 \times 10^{-6}  \;\mu_{B}. \ee

This bound is in good agreement with that found by the L3 and
DELPHI Collaborations  \cite{L3-1}. It compares favorably with the
limits $\mu_{\nu{\tau}}<4\times 10^{-6} \; \mu_B$~ \cite{Grotch}
and $\mu_{\nu_{\tau}}<2.7\times 10^{-6} \; \mu_B$~
\cite{Escribano} obtained from low--energy experiments and from
the invisible width of the $Z$ boson, respectively. However, all
these bounds are still weaker than the experimental bounds
$\mu_{\nu_e}<1.1\times 10^{-10} \; \mu_B$,
$\mu_{\nu_\mu}<7.4\times 10^{-9} \;\mu_B$  \cite{Krakauer}, and
$\mu_{\nu_\tau}<5.4\times 10^{-7} \;\mu_B$  \cite{Cooper}, or the
most stringent bound $(0.2-0.8)\times 10^{-11} \;\mu_B$ obtained
from chirality flip in the 1987 Supernova explosion and valid for
the three neutrino flavors  \cite{Ayala}.

The LEP--1 limit (\ref{BRZnng}) on $Z \to \bar\nu\nu\gamma$ can
also be used to get bounds on the effective coupling
$Z\bar\nu\nu\gamma$, which is generated by the dimension--six
operators  \cite{Dusedau,Maya}

\be {\cal L}_{Eff.}=\frac{\alpha_1}{\Lambda^2}{\bar
\ell}^a_L\tau^i\gamma^\mu D^\nu \ell^a_L W^i_{\mu\nu}+
\frac{\alpha_2}{\Lambda^2}{\bar \ell}^a_L\tau^i\gamma^\mu D^\nu
\ell^a_L B_{\mu\nu}, \label{LZnng} \ee

\noindent where ${\bar \ell}^a_L$ is the left--handed doublet,
$W^i_{\mu\nu}$ and $B_{\mu\nu}$ are the $SU_L(2)$ and $U_{Y}(1)$
strength tensors, respectively, and $D^\mu$ is the covariant
derivative. The bound obtained for the coefficients of these
operators is given by $\epsilon_8<0.165$  \cite{Maya}, with the
following definition for this dimensionless coupling

\be
\epsilon_8=(\alpha_1+\alpha_2)\left(\frac{v}{\Lambda^2}\right).\ee

In the SM, the dimension--six operators given in (\ref{LZnng}) may
be induced by the diagrams shown in Fig.
\ref{ztnngSMfd}d--\ref{ztnngSMfd}j. However, this contribution is
negligibly small, of the order of $10^{-8}$  \cite{Hernandez-1}.
In the strongly--coupled standard model they are induced by
similar box diagrams with excited vector bosons or leptoquarks
 \cite{Dusedau}. In this case, the limit $\epsilon_8<0.165$ can be
used to set bounds on the masses or couplings of these new degrees
of freedom. However, this calculation has not been done to our
knowledge.

\section{Two--photon decays}

Since the decay $Z\to \gamma\gamma$ is forbidden by Bose symmetry
and angular momentum conservation  \cite{Landau}, two--photon
events may arise from the decay of the $Z$ boson into a photon
pair plus a neutrino pair $Z\to \bar\nu\nu\gamma\gamma$, which has
been studied in order to constrain new physics effects. The L3 and
OPAL Collaborations have looked for events with a photon pair of
large invariant mass accompanied by a lepton pair and have put the
bound  \cite{L3-2}

\be \label{BRZnngg}BR(Z\to \bar\nu\nu \gamma\gamma) < 3.1\times
10^{-6}. \ee

\begin{figure}
\begin{center}
\epsfig{file=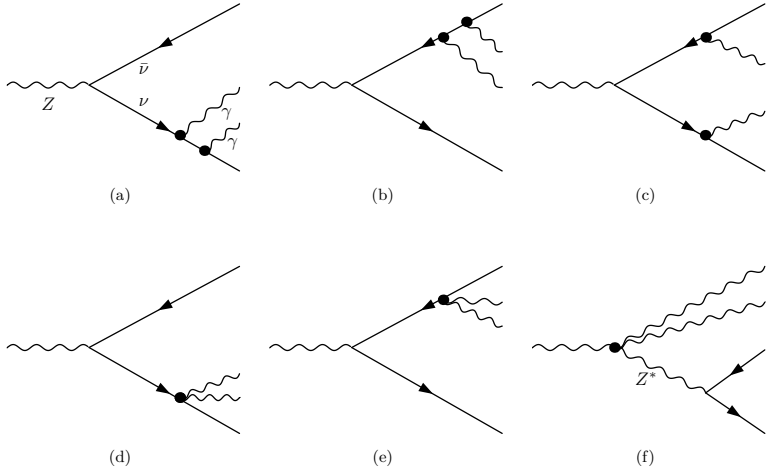,width=4in} \caption{Feynman diagrams
contributing to the decay $\decb$ in the effective Lagrangian
scheme {\protect \cite{Larios-2}}. Crossed diagrams are not
shown.} \label{ztnnggELfd}
\end{center}
\end{figure}

The respective decay width for this mode has not been computed in
the SM to our knowledge. It is expected to be suppressed with
respect to the $Z\to \nu\bar\nu\gamma$ decay width
 \cite{Hernandez-1} by an additional $\alpha$ factor. In the
effective Lagrangian formalism, the two--photon decay mode is
generated by the Feynman diagrams shown in Fig. \ref{ztnnggELfd}
 \cite{Larios-2,Perez-1}. Besides the contributions induced by the
neutrino magnetic dipole transition $\nu\bar\nu\gamma$ given in
Fig. \ref{ztnnggELfd}a--\ref{ztnnggELfd}c, it is necessary to
include the contributions associated with the
neutrino--two--photon interaction $\bar\nu\nu\gamma\gamma$ (Fig.
\ref{ztnnggELfd}d--\ref{ztnnggELfd}e) and the quartic gauge boson
coupling $ZZ\gamma\gamma$ (Fig. \ref{ztnnggELfd}f).

Neutrino--two--photon interactions may have direct implications on
some astrophysical processes such as the cooling of stars by a
high annihilation rate of photons into neutrinos  \cite{Levine}.
This interaction can be parametrized with the following effective
Lagrangian  \cite{Nieves}

\be {\cal L}^{\bar \nu_i\nu_j \gamma \gamma}_{Eff.}=\frac{1}{4
\,\Lambda^3} \,\bar \nu_i\,\left(\alpha^{ij}_L P_L+\alpha^{ij}_R
P_R \right) \,\nu_j \tilde{F}^{\mu\nu} F_{\mu\nu} ,\label{Lnngg}
\ee

\noindent where $\alpha_{ij}$ are dimensionless coupling
constants. This interaction induces the following contribution to
the $Z\to \bar\nu\nu\gamma\gamma$ branching ratio  \cite{Larios-2}

\be BR(\decb)=1.092\times 10^3\,
\sum_i\sum_j\left(|\alpha^{ij}_L|^2+|\alpha^{ij}_R|^2\right)\frac{1}{\Lambda^6},
\ee

\noindent where the sum runs over all neutrino species and
$\Lambda$ should be expressed in GeV. The experimental bound given
in (\ref{BRZnngg}) induces the limit

\be \frac{1}{\Lambda^6}
\sum_i\sum_j\left(|\alpha^{ij}_L|^2+|\alpha^{ij}_R|^2\right)\leq
2.85 \times 10^{-9}. \label{bound3} \ee

\noindent which in turn can be translated into a lower bound on
the lifetime of the neutrino double radiative decay $\nu_i\to
\nu_j \gamma\gamma$  \cite{Larios-2}

\be \tau_{\nu_{i}}\ge  1.79\times10^{12}\Bigg[\frac{1\,\mathrm{
MeV}}{m_{\nu_j}}\Bigg]^7\;\, \mathrm{s}. \label{ltbound}\ee

This limit is one order of magnitude weaker than that obtained
from the analysis of the Primakoff effect on the process $\nu + N
\to \nu + N$ in the presence of the external field of a nucleus
$N$  \cite{Gnienko}. Nevertheless, as occurred with the bounds
obtained from the $Z\to \bar\nu\nu\gamma$ mode, the advantage of
the bound (\ref{ltbound}) for the neutrino--two--photon
interaction is that it is a model--independent result and relies
on very few assumptions.

The limit (\ref{BRZnngg}) on the decay width of $Z\to
\bar\nu\nu\gamma\gamma$ can also be used to constrain the quartic
neutral gauge boson (QNGB) coupling shown in Fig. \ref{ztnnggELfd}
f  \cite{Perez-1} An interesting feature of the QNGB couplings
that involve at least one photon field is that they are induced by
effective operators which are not related to the triple neutral
gauge boson couplings $V_iV_jV_k$. As a consequence, the known
constraints on the latter do not apply to the former and it is
thus necessary to study them in an independent way. The lowest
dimension operators that induce the $ZZ\gamma\gamma$ coupling have
dimension six (eight) in the nonlinear (linear) realization of
effective Lagrangians. New physics effects could become more
evident thus in the nonlinear scenario. In such case, there are
fourteen dimension--six operators that induce the $ZZ\gamma\gamma$
effective coupling  \cite{Belanguer}. However, in the unitary
gauge there are only two independent Lorentz structures for this
coupling

\be \label{eflag} {\cal L}^{ZZ\gamma
\gamma}_{Eff.}=-\frac{e^2}{16\, \Lambda^2\,c_W^2}a_0 F_{\mu
\nu}F^{\mu \nu}Z^\alpha
Z_\alpha-\frac{e^2}{16\,\Lambda^2\,c_W^2}a_c F_{\mu \nu}F^{\mu
\alpha}Z^\nu Z_\alpha, \ee

\noindent which in turn give the following contribution to the
decay width  \cite{Perez-1}

\be \Gamma(Z\to \bar\nu \nu\gamma\gamma)=\left(\frac{1\;{\rm{GeV}}
}{\Lambda}\right)^4
\left(N_{0c}\,|4\,a_0+a_c|^2+N_{c}\,|a_c|^2\right)\;\;{\rm{GeV}},
\label{result} \ee

\noindent with $N_{0c}\approx 3.46\times 10^{-6}$ and
$N_{c}\approx 10.31\times 10^{-6}$. If one assumes that either
$a_o$ or $a_c$ is dominant, then the following bounds are obtained
 \cite{Perez-1}

\bae
\frac{|a_0|}{\Lambda^2}&\leq& 0.106 \;\;{\rm GeV}^{-2}\quad{\rm if}\quad a_0\gg a_c,\nonumber\\
\frac{|a_c|}{\Lambda^2}&\leq& 0.215\;\;{\rm GeV}^{-2}\quad{\rm
if}\quad a_c\gg a_0.
\eae

These limits are weaker by about one order of magnitude than those
obtained at LEP--2 from $Z\gamma\gamma$ and $WW\gamma$ production
 \cite{L3-3}.

\section{Decays with photons and gluons}

As already mentioned, the decay of the $Z$ boson into two massless
vector particles ($Z\to\gamma\gamma$, $gg$) is forbidden by the
Landau--Yang theorem  \cite{Landau}, while the decay $Z\to \gamma
g$ is also forbidden by color conservation. On the other hand, the
rare decays $Z\to \gamma\gamma\gamma$, $ggg$, $\gamma gg$ can be
induced in the SM only at the loop level: the coupling of the $Z$
boson to three gauge vector particles requires an effective
interaction of dimension higher than four with three tensor fields
$F_{\mu\nu}$ or $\tilde{F}_{\mu\nu}$.

In the case of the three--photon decay mode, the fermion
 \cite{Laursen-1} and the vector boson  \cite{Baillargeas-1}
contributions are of the order of $10^{-10}$ and $10^{-11}$,
respectively, whereas the charged scalar boson contribution is
about four orders below  \cite{Konig}. The branching ratios for
decays involving gluons are somewhat higher
 \cite{Laursen-1,Laursen-2,Lee}: $BR(Z\to ggg)\sim 1.8\times
10^{-5}$ and $BR(Z\to \gamma gg)\sim 4.9\times 10^{-6}$. In spite
of the smallness of these branching ratios, there has been some
interest in estimating new physics effects in the rare $Z$ decay
modes involving three vector gauge bosons
 \cite{Stohr,Baillargeon-2}. Furthermore, since the decay amplitude
for the three--photon mode is proportional to the cubic of the
electric charge of the particles circulating in the box diagrams,
it might be possible that the contribution of particles with
charge greater than unity, such as doubly charged ones, may induce
a dramatic enhancement similar to that expected in $\gamma\gamma$
collisions  \cite{Tavares}.

In the ELA there are two independent operators of dimension eight,
which are $U(1)$ invariant and CP conserving, inducing the
$Z\gamma\gamma\gamma$ coupling  \cite{Stohr,Baillargeon-2}

\be {\cal L}_{Eff.}^{Z\gamma\gamma\gamma}=G_1\,F^{\alpha
\mu}F^{\sigma\nu}\partial_\alpha F_{\mu\nu} Z_\sigma
+G_2\,F^{\alpha \beta}F^\nu_\beta\partial_\alpha F_{\sigma\nu}
Z^\sigma,\ee

\noindent This interaction induces a decay width given by

\be
\Gamma(Z\to\gamma\gamma\gamma)=\frac{m_Z^9}{55960\pi^3}\left(2G_1^2+3G_2^2-3G_1G_2\right).
\ee

Of course, this general result can be used to get bounds on the
$G_1$ and $G_2$ coupling constants once we have a sensible limit
on this decay mode.

The possibility of inducing the two--gauge--bosons decay modes
$Z\to \gamma\gamma$, $gg$ has been explored in a background
magnetic field  \cite{Tinsley}. In principle, these decay modes
should be equivalent to the $Z\to \gamma\gamma\gamma$ and $\gamma
gg$ decay channels in vacuum. Accordingly, the respective
branching ratios come out of the same order of magnitude
 \cite{Tinsley}: $BR(Z\to \gamma\gamma)\sim 10^{-11} (B/B_o)^2$,
$BR(Z\to gg)\sim 10^{-10}(B/B_o)^2$, where $B$ is the strength of
the background magnetic field and $B_o=m_e^2/e$.

\section{Decays involving scalars}

Until now, the only missing ingredient of the SM is the Higgs
scalar boson. Even more, in many beyond--the--SM extensions there
is the prediction of more than one Higgs boson  \cite{HHunter}.
For instance, the simplest extension of the SM is comprised by two
Higgs scalar doublets and predicts five scalar bosons: two
CP--even scalar bosons $h$ and $H$, one $CP$--odd scalar boson
$A$, and one pair of charged scalar bosons $H^\pm$. There has been
thus a great deal of interest in studying Higgs boson production
from $Z$ decays. Currently, the nonobservation of $e^-e^+\to
Z^*\to Z h$ has set the bound $m_h>114.7$ GeV on the SM Higgs
boson mass  \cite{Talks}. However, it has been argued that there
are some theories in which the $ZZh$ coupling may be largely
suppressed, thereby weakening the above bound to a great extent
 \cite{Dedes,Duong-1}. It is thus possible that a light Higgs
boson, with a mass $m_h<m_Z/2$, may have escaped detection so far
via Higgs boson radiation off a $Z$ boson at LEP--2. On the other
hand, it has also been pointed out that the $ZZhh$ coupling
happens to be model independent and unsuppressed  \cite{Duong-1}.
In this scenario, there is the chance that some rare decays of the
$Z$ boson into two or three Higgs bosons may be kinematically
allowed and at the reach of future colliders
 \cite{Duong-1,Duong-2}. Since Bose symmetry forbids the $Z\to hh$
decay, other modes have to be studied. Among them, the decay modes
$Z\to h_5^0\chi^0\chi^0$ and $Z\to h_5^0 h_5^0 h_5^0$, with
$h^0_5$ a very light scalar boson and $\chi^0$ a massless Majoron,
have been studied in the framework of a doublet Majoron model
 \cite{Duong-2}. It was found that the respective branching ratio
may reach the level of $10^{-7}$, which is also valid for models
with a more exotic Higgs sector. The existence of a very light
Higgs scalar would kinematically allow also the rare $Z\to hh\bar
f f$ decay, with $f$ a light fermion, which may occur with a
branching ratio of the order of $10^{-7}$. This decay mode would
be observable especially in models in which the scalar boson $h$
decays invisibly, such as in some Majoron models  \cite{Duong-1}.

As far as the CP--odd Higgs boson $A$ is concerned, the current
bounds on its mass are model dependent and a light CP--odd scalar
is still not ruled out in some specific models
 \cite{Larios-3,Sher}. Even more, some SM extensions, such as the
minimal composite Higgs model \cite{Dobrescu-1} or the
next--to--minimal supersymmetric standard model  \cite{HHunter},
do predict a very light CP--odd scalar. Even if such a light
particle happens to exist, the rare decay $Z\to hA$ \cite{Djouadi}
would not be kinematically allowed for a CP--even Higgs boson
whose mass is close to the current lower bound $m_h>114.7$ GeV.
Nevertheless, it is still feasible to look for a light $A$ as the
product of other rare $Z$ decays such as $Z\to A\,A\,\bar l\,l$
\cite{Haber}. A situation resembling that discussed above for the
CP--even scalar $h$ arises for the CP--odd scalar: while the $ZZA$
coupling is absent at the tree level, the $ZZAA$ coupling is fixed
by the gauge invariance of the theory and it is not suppressed. As
a consequence, the decay mode $Z\to AA\,\bar l l$ may be feasible
for a light CP--odd scalar, with the lepton pair arising from a
virtual $Z$ boson, even though with a small branching ratio of the
order of $10^{-8}$  \cite{Haber}.

\begin{figure}
\begin{tabular}{cc}
\epsfig{file=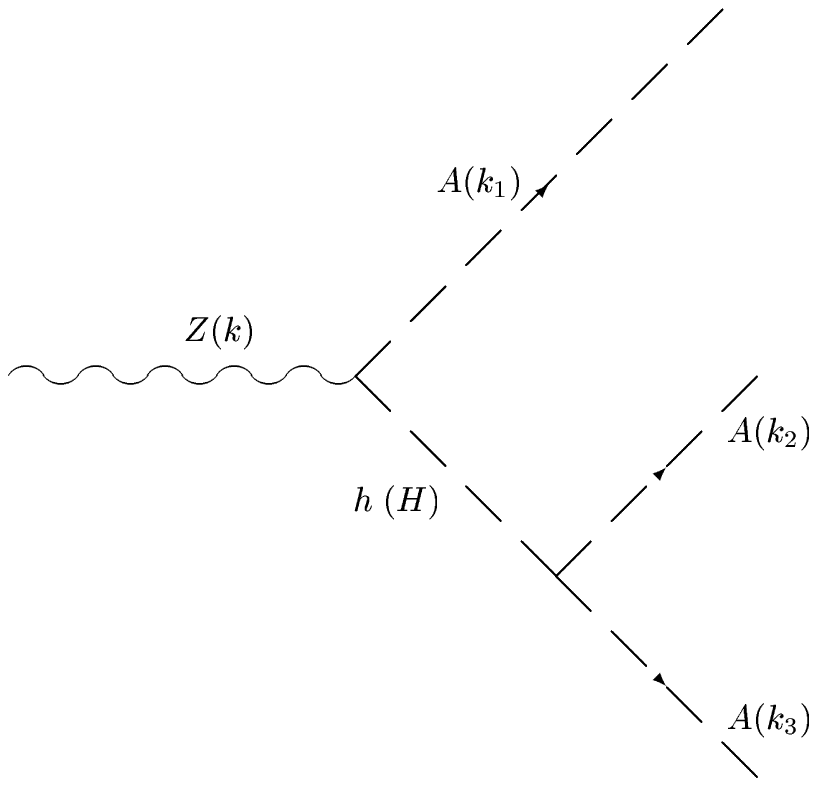,width=1.7in,height=1.5in}&
\epsfig{file=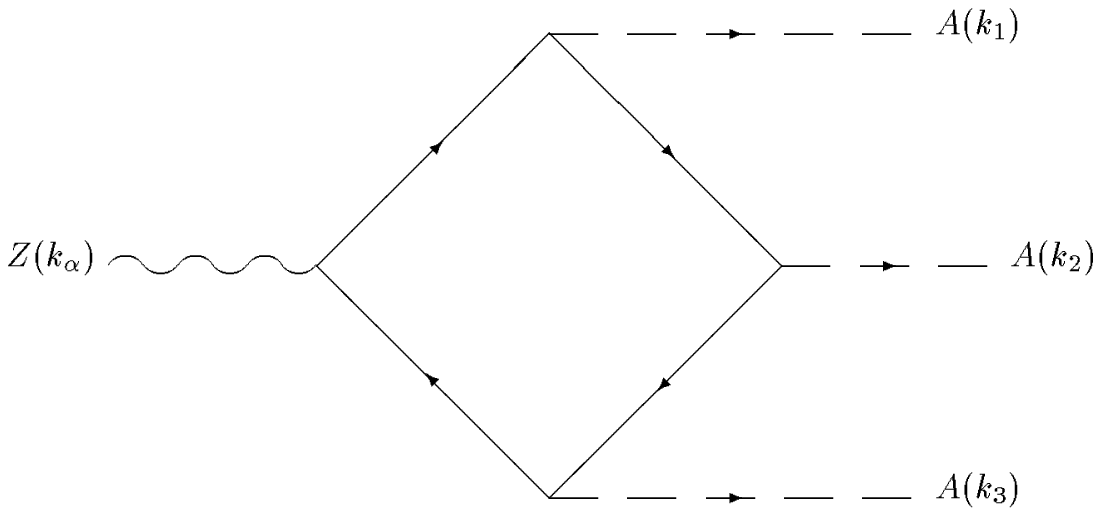,width=3in,height=1.5in}\\
{\tiny (a)}&{\tiny (b)}
\end{tabular}
\caption{Feynman diagrams contributing to the decay $Z\to AAA$ at
the tree and one--loop level in the THDM. Charged fermions
circulate in the loop.} \label{ZtoAAA}
\end{figure}

If $A$ is very light, the $Z\to AAA$ decay mode would be
kinematically allowed  \cite{Larios-3,Chang,Ma}. In the
two--Higgs--doublet model (THDM), this decay may proceed at the
tree level by the exchange of the CP even Higgs bosons $h$ and $H$
as depicted in Fig. \ref{ZtoAAA}a. The three level induced $\phi
AA$ coupling, with $\phi=h\; (H)$, can be written as  \cite{Chang}

\be {\cal L}_{Eff.}^{\phi AA}=\lambda \, \phi A A \label{hAAc},
\ee

\noindent where $\lambda$ lies in the Fermi scale in several
specific models  \cite{Chang}. The contribution of a CP--even
Higgs boson $h$ to the rare $Z\to AAA$ decay is

\begin{equation}
{\cal M}(Z \to AAA)=\frac{2 g \, \lambda}{\cos
\theta_W}\sum_{i=1}^3 \frac{\epsilon(k,\lambda) \cdot
k_i}{(k-k_i)^2-m_h^2}. \label{mtlzaaa}
\end{equation}

\noindent Assuming a typical value of $\lambda\sim 100$ GeV, it
was found that $BR(Z\to AAA)\sim 10^{-5}$ when $m_h\sim m_Z$ and
$m_A\ll m_Z$  \cite{Chang,Ma}. This branching fraction drops
suddenly as $m_h$ becomes heavier than $m_Z$ and $m_A$ approaches
$m_Z/3$, the upper limit allowed by the kinematics of the process.
There is also the possibility of a large contribution to $Z\to
AAA$ coming from loop diagrams  \cite{Larios-3,Chang}. At the
one--loop level (Fig. \ref{ZtoAAA}b), this decay is induced by
fermion loops whose contribution to the decay width is
proportional to $m_f^6\,C_f^6$, where $C_f$ is the strength of the
$A \bar f f$ coupling.  It might be that $C_f$ were so large that
this enhancement factor would overcome the natural suppression
factor coming from the loop. In this case the dominant
contributions are those from the $b$ and $t$ quarks. The effective
$A\bar q q$ coupling can be written as

\be {\cal L}_{Eff.}^{A\bar q q}=-\frac{g}{2 m_W}\sum_{q} m_q C_q
\bar{u}_q \gamma^5 u_q \ A, \ee

In THDMs type I, $C_q=\tan \beta\,(\cot\beta)$ for up (down)
quarks, whereas in THDMs type II $C_q=\cot \beta$ for any quark.
The one--loop contribution was roughly estimated in Ref.
 \cite{Chang} and the exact calculation was presented in the
appendix of Ref.  \cite{Larios-3}. In the $m_A\to 0$ limit, it was
found that

\begin{equation}
\label{brZtoAAA} {BR}(Z \to AAA)=1.3 \times 10^{-18} C_t^6+2.47
\times 10^{-17} C_b^6 + 7.63 \times 10^{-18} C_t^3 C_b^3.
\end{equation}

\noindent It turns out that the $b$ contribution is larger than
that of the $t$ quark as long as $C_b>C_t$. However, because of
unitarity, $C_b$ cannot be arbitrarily large. Furthermore,
requiring the validity of perturbation theory yields the bound
$C_b<120$. In this limit, $BR(Z\to AAA)\sim 10^{-5}$
 \cite{Larios-3}. To asses the possibility of observing this decay
mode, a more realistic analysis is indeed required. In particular,
the parameters of the THDM should be constrained from the current
low--energy data on several observables, such as the $\rho$
parameter, $BR(b\to s\gamma$), $R_b$, $A_b$, $BR(\Psi\to A\gamma$)
and $(g-2)$ of $\mu$. Such an analysis was presented in Refs.
 \cite{Larios-3},  \cite{Sher} and  \cite{Cheung}. It was found that
the low--energy data still leave open a small window for the
existence of a CP--odd scalar $A$ with a mass as light as
$m_A<0.2$ GeV  \cite{Larios-3}. Unfortunately, the remaining
parameters of the model are tightly constrained. In this scenario,
the triple pseudoscalar decay $Z\to AAA$ may occur with a
branching ratio of the order of $10^{-8}$
 \cite{Larios-3,Chang,Ma}, arising mainly from the three--level
contribution. The signature of this decay would be spectacular:
each one of the three CP--odd scalars $A$ will decay predominately
into a photon pair, which in turn will be registered in the
detectors of high energy colliders as a single photon when the
momentum of $A$ is much larger than its mass  \cite{Dobrescu-2}.

Finally, as far as the charged Higgs scalar is concerned, it seems
that there is no way that this particle can be produced from $Z$
decays according to the current bounds on $m_{H^\pm}$, obtained
from direct searches at LEP--2 and indirectly from the analysis of
the bounds on low--energy observables  \cite{PDG,Gambino}.

\section{Flavor changing decays}

Since lepton flavor violation (LFV) is forbidden in the SM, the
rare $Z\to l_i^\mp l_j^\pm$ decays, with $l_i=e$, $\mu$ and
$\tau$, have been widely studied as the detection of any effect of
this kind would serve as an indisputable evidence of new physics.
Furthermore, the possibility that neutrinos have nonzero mass
 \cite{SK,SNO,LSND}, which in turn would allow lepton flavor
transitions, has boosted considerably the interest in LFV
processes. Even if the SM were enlarged with massive Dirac or
Majorana neutrinos, the rate for $Z\to l_i^\mp l_j^\pm$ would be
of the order of $10^{-54}$  \cite{Illana-1}, which means that this
class of decays would be very sensitive to new physics effects
induced by other SM extensions. The current experimental bounds on
the LFV $Z$ decays were obtained at the CERN LEP--1 collider
 \cite{PDG}

\bae BR\left(Z \to e^{\mp} \mu^{\pm}\right)&<& 1.7 \times
10^{-6},\nonumber\\
BR\left(Z \to e^{\mp} \tau^{\pm}\right)&<& 9.8 \times
10^{-6},\nonumber\\
BR\left(Z \to\mu^{\mp} \tau^{\pm}\right)&<&1.2 \times 10^{-5}.
\eae

A plenty of work has been done in the past to analyze LFV $Z$
decays, which have been approached in two different ways:
model--independent analyses and predictions from specific
extensions of the SM. In the former case, the starting point is
the effective Lagrangian which leads to the most general structure
for the $Zl_il_j$ effective vertex  \cite{Nussinov,Flores}

\be
\label{MZlilj}
 {\cal M}^{Zl_il_j}=\frac{ig}{2c_W} \bar{u}(p_i)\left(
\gamma_\mu\left(F_{1L}^{ij} P_L + F_{1R}^{ij} P_R\right) +
\frac{i}{m_Z} F_{3R}^{ij} P_R \sigma_{\mu \nu} k^\nu \right)v(p_j)
Z^\mu, \ee

\noindent We have dropped the $k_\mu$ term from Eq. (\ref{MZlilj})
as it does not contribute when the $Z$ boson is on its
mass--shell. In the ELA, the monopole and dipole moment
contributions can be generated by the following effective
operators  \cite{Flores}

\bae \label{monoZlilj} {\cal O}_{\phi \ell}^{ij}&=&i\left(
\phi^{\dagger }D_\mu \phi \right) \left( \bar \ell_{Ri}\gamma^\mu
\ell_{Rj}\right),\nonumber\\
{\cal O}_{\phi L}^{(1)ij}&=&i\left( \phi^{\dagger
}D_\mu \phi \right) \left( \bar L_i\gamma ^\mu L_j\right),\nonumber\\
 {\cal O}_{\phi L}^{(3)ij}&=&i\left( \phi^{\dagger
}\tau^a D_\mu \phi \right) \left( \bar L_i\tau ^a\gamma^\mu
L_j\right),
\eae

\bae \label{dipZlilj} {\cal O}_{\ell W\phi}^{ij}&=& g \left(\bar
L_i \sigma_{\mu \nu} {\bf W}^{\mu \nu} \ell_{R j}\right)
\phi,\nonumber\\
{\cal O}_{\ell B\phi}^{ij}&=& g^\prime \left(\bar L_i \sigma_{\mu
\nu} B^{\mu \nu} \ell_{R j}\right) \phi. \eae

\noindent where $L_i$ and $\ell_{Ri}$ stand for the left--handed
doublet and right--handed singlet of the $SU(2)\times U(1)$ gauge
group, respectively. The monopole moment structures $F_{1\,L,\,R}$
arise from the operators (\ref{monoZlilj}), which in turn are
generated at the tree level in the underlying theory, while the
dipole moment structure $F_{3R}$ is induced by the operators
(\ref{dipZlilj}). The latter can arise only at the one--loop level
in the underlying theory \cite{Artz} and their contribution has an
additional suppression factor of the order of $(4\pi)^{-1}$. It is
thus a good approximation to consider only the contribution
arising from the operators (\ref{monoZlilj}):

\be \label{BRZlilj} BR\left(Z\to l_i^{\mp}l_j^{\pm}\right) =
\frac{\alpha}{3 s_{2W}^2} \left(\frac{m_Z}{\Gamma_Z}\right) \left(
|F_{1L}^{ij}|^2 + |F_{1R}^{ij}|^2 \right). \ee

The effective coupling given in Eq.(\ref{MZlilj}) induces also a
contribution to the LFV decays $l_i\to l_j \bar l_k l_k$. Using
the known experimental limits on $BR(l_i\to l_j \bar l_k l_k$)
 \cite{PDG}, the following bounds on LFV $Z$ decays are thus
obtained  \cite{Nussinov,Flores}

\bae \label{Zliljbound} BR\left(Z \to \mu^{\mp} e^{\pm}\right)
&\le&5.0\times 10^{-13}\,(1.04
\times 10^{-12}),\nonumber\\
BR\left(Z \to \tau^{\mp} e^{\pm}\right)  &\le&3.0\times 10^{-6}\,
(1.7
\times 10^{-5}),\nonumber\\
BR\left(Z \to \tau^{\mp} \mu^{\pm}\right) &\le&3.0\times 10^{-6}\,
(1.0 \times 10^{-5}), \eae

\noindent where the quantities given in the right--hand side of
(\ref{Zliljbound}) corres\-pond to the limits obtained from
unitarity--inspired arguments  \cite{Nussinov}, while the ones in
parenthesis are obtained from the analysis performed in the
effective Lagrangian approach  \cite{Flores}. Along this line, it
is interesting to note that the analysis of $\mu\to e$ conversion
in a nuclear field leads to indirect bounds on the branching
ratios of the LFV decay $Z\to l_i^\mp l_j^\pm$, which are in
agreement with those shown in Eq. (\ref{Zliljbound})
 \cite{Delepine}.

LFV decay modes of the $Z$ boson have also been studied in a wide
variety of extensions of the SM, from which we would like to
mention some of the more representative: left--right symmetric
models  \cite{Perez-2}, SUSY  \cite{Illana-2}, left--right
supersymmetric models  \cite{Frank}, the Zee model  \cite{Ghosal},
theories with a heavy $Z'$ boson  \cite{Langaker}, the SM enlarged
with massive neutrinos,  \cite{Illana-1}, universal top color
assisted technicolor models  \cite{Yue}, and the general two Higgs
doublet model (THDM--III)  \cite{Iltan}. For a more comprehensive
list of the literature dealing with this topic, the reader is
referred to Ref.  \cite{Eilam}. It turns out that one of the more
promising scenarios for LFV $Z$ decays is that favored by the
type--III THDM, in which the $Z\to l_i l_j$ decay mode proceeds at
the one--loop level via the exchange of a virtual Higgs boson with
tree-level LFV couplings $H\, l_i\,l_j$  \cite{Diaz-2}. This model
predicts a branching ratio large enough to be tested within the
expected sensitivity of the giga--$Z$ linear collider
 \cite{Aguilar}. In this case, it was found that $BR(Z\to e\mu)\sim
10^{-11}$ and $BR(Z\to \tau e,\tau \mu) \sim 10^{-10}$
 \cite{Iltan}.

Although FCNC effects in the quark sector have been extensively
studied through the processes $b\to s \gamma$  \cite{Campbell},
and $t\to c\gamma$  \cite{Diaz}, rare FCNC $Z$ decays are also
good candidates to look for any new physics effects and have been
the source of great interest recently. Although this class of
transitions are only forbidden at the three level and can arise at
the one--loop level, the GIM mechanism suppresses them
effectively. In the SM, the one--loop induced FCNC $Z q q^\prime$
coupling was calculated first in the context of the $K_L \to \mu
\bar\mu$ decay and in the limit of massless external quarks
 \cite{Gaillard}. The calculation was later generalized to the case
with massive internal and external up and down quarks
 \cite{axelrod}. Afterwards, the effects of a fourth fermion family
 \cite{Ganapathi} and the possibility of CP violation  \cite{Hou} in
the $Z\to qq^\prime$ decay were also examined. As for the dominant
decay channel $Z\to b \bar s$\footnote{Unless stated otherwise,
$Z\to b\bar s$ stands for $Z\to b\bar s+\bar b s$.}, the
respective branching ratio is $BR(Z\to b\bar s)\sim 3\times
10^{-8}$ in the SM  \cite{axelrod}. This decay mode has been
studied also in several extensions of the SM: THDM type II
\cite{Bush} and type III  \cite{Atwood-1}, SUSY models
 \cite{Mukhopadhyaya}, and SUSY models with broken $R$--parity
 \cite{Chemtob}. The predictions for $BR(Z\to b \bar s)$ in these
models happen to be very small and this rare $Z$ decay seems
beyond the reach of the future colliders. However, quite recently,
various scenarios have been considered in SUSY models with flavor
violation in the scalar sector  \cite{Atwood-2}. In this case it
was found that $BR(Z\to b\bar s)$ can reach $10^{-6}$ in SUSY
models with mixing between the bottom and strange--type squarks
and/or mixings between sleptons and Higgs fields for large
$\tan\beta$ values. A similar conclusion was reached in the
context of topcolor--assisted technicolor models  \cite{Hue},
where it was found that the contribution coming from top--pions
can reach $BR(Z\to b \bar s)\sim 10^{-5}$. For other works on the
rare $Z\to qq^\prime$ decay, we refer the reader to Ref.
 \cite{Park}.

It is interesting to notice that any FCNC effect  at the level of
$BR(Z\to l_il_j)\sim 10^{-10}$-- $10^{-8}$ or $BR(Z\to bs)\sim
10^{-7}$--$10^{-6}$ would be at the reach of the expected
sensitivity of the giga--$Z$ linear collider  \cite{Aguilar}.
While the prediction for LFV $Z$ decays is expected to reach the
giga--$Z$ experimental upper limit  \cite{Iltan} in models such as
the THDM type III, with $BR(Z\to l_i^\mp l_j^\pm)\sim
10^{-11}$--$10^{-10}$, one can get at most $BR(Z\to b \bar s)\sim
10^{-8}$ in the same model with the current experimental
constraints, which in turn will be out of the reach of the
giga--$Z$ linear collider  \cite{Atwood-1,Atwood-2}.

\section{Concluding remarks}

The study of virtual effects induced by new physics in rare $Z$
decays provides an important opportunity to probe the presence of
interactions beyond the SM. In the present review we have
appreciated that there is a complementary approach between the
results obtained within the framework of radiative corrections to
perturbatively calculable processes in the SM and transitions
which are either suppressed or forbidden in the SM. The more
promising situations arise when SM predictions are well below the
expectations coming from new physics effects. A summary of the
rare $Z$ decay modes considered in this article is presented in
Table 1. We have included the existing experimental bounds, the
respective SM predictions and the main new physics effects that
may be tested with the expected sensitivity of the giga--$Z$
linear collider  \cite{Aguilar}.

\begin{table}[!hbt]
\caption{Summary of rare $Z$ decay modes. References to specific
results appear in brackets. $\Delta^{\pm\pm}$ stands for a doubly
charged particle. We only show those new physics effects that
appear to be at the reach of the giga--$Z$ linear collider.}
\begin{center}
{\begin{tabular}{@{}cccl@{}}\hline
Decay mode&Experimental bound (BR)&SM prediction (BR)&New Physics effects\\
\hline $\deca$&$1.0\times 10^{-6}$~ \cite{L3-1}&
$7.1\times 10^{-10}$~ \cite{Hernandez-1}&$ZZ\gamma$~ \cite{Larios-1}\\
&&&$\bar\nu\nu\gamma$~ \cite{Maya,Maltoni,Aydin}\\
&&&$\bar\nu\nu\gamma Z$~ \cite{Maya}\\
&&&Light $\widetilde G$, $\widetilde Z$~ \cite{Dicus}\\&&&Light $J$~ \cite{Romao-1,Romao-2}\\
\hline
$\decb$&$3.1\times 10^{-8}$~ \cite{L3-2}&&$ZZ\gamma\gamma$~ \cite{Perez-1}\\
&&&$\bar\nu\nu\gamma\gamma$~ \cite{Larios-2} \\
\hline $Z\to\gamma\gamma\gamma$&$1.3\times 10^{-5}$ ~
\cite{DELPHI-1}&$1.0\times 10^{-10}$~
\cite{Laursen-1,Baillargeas-1}&Light $A$
~ \cite{Kim,Larios-3,Chang}\\
$Z\to ggg$&$1.8\times10^{-2}$~ \cite{DELPHI-2}&$1.8\times
10^{-5}$~ \cite{Laursen-2}&$\Delta^{\pm\pm}$~ \cite{Tavares}\\
$Z\to g g\gamma$&&$4.9\times 10^{-6}$~ \cite{Laursen-2}&$ZV_iV_jV_k$\\
\hline $Z \to AAA$&&&Light $A$ ~ \cite{Kim,Larios-3,Chang}\\
\hline
$Z\to e^\pm \mu^\mp$&$1.7\times 10^{-6}$~ \cite{PDG}&0&THDM--III~ \cite{Iltan}\\
$Z\to e^\pm \tau^\mp$&$9.8\times 10^{-6}$~ \cite{PDG}&0&\\
$Z\to \mu^\pm \tau^\mp$&$1.2\times 10^{-5}$~ \cite{PDG}&0&\\
\hline $Z\to b\bar s$&&$3.0\times 10^{-8}$&SUSY~ \cite{Atwood-2}\\
&&&Technicolor~ \cite{Hue}\\
\hline
\end{tabular}}
\end{center}
\end{table}

We would like to close by stating that even in case that no new
physics effects were discovered in the planned giga--$Z$ linear
collider, an improvement in the known experimental bounds on these
processes will still provide a critical test of the validity of
the SM at the loop level.

\section*{Acknowledgments}

We would like to thank discussions with E. Ma, C.-P. Yuan and F.
Larios. Support from CONACyT and SNI (Mexico) is also
acknowledged. The work of G. T. V. is also supported by
SEP-PROMEP.

\end{document}